\documentclass[onecolumn,authoryear]{els-mrw} 

\usepackage{amsmath,amssymb,amsfonts,amsthm,makeidx,graphicx}
\usepackage{txfonts}
\usepackage{helvet}
\usepackage{aas_macros}
\usepackage{listings}
\usepackage{hyperref}
\usepackage{xcolor}
\hypersetup{colorlinks = True}
\usepackage{enumitem}
\usepackage{tcolorbox}
\usepackage{totcount}
\newtotcounter{citnum} 
\def\oldbibitem{} \let\oldbibitem=\bibitem
\def\bibitem{\stepcounter{citnum}\oldbibitem}

\begin{document}
\chapter{Main-sequence systems: orbital stability around single star hosts}\label{chap1}

\author[1,2]{Hareesh Gautham Bhaskar}%
\author[1]{Nathaniel W. H. Moore}%
\author[1]{Jiapeng Gao}%
\author[1]{Gongjie Li}%
\author[3]{Billy Quarles}%

\address[1]{\orgname{Georgia Institute of Technology}, \orgdiv{School of Physics}, \orgaddress{Howey Physics Bldg, 837 State St NW, Atlanta, GA 30332}}
\address[2]{\orgname{Technion - Israel Institute of Technology}, \orgdiv{Department of Physics}, \orgaddress{Haifa 3200003 Israel}}
\address[3]{\orgname{Texas A\&M University-Commerce}, \orgdiv{Department of Physics \& Astronomy}, \orgaddress{P.O. Box 3011 Commerce, TX 75429-3011, USA}}

\articletag{Chapter Article tagline: update of previous edition,, reprint..}

\maketitle





\begin{abstract}[Abstract]
Stability is one of the most fundamental aspects regarding planetary systems. It plays an important role in our understanding on the formation channel of the planetary systems, as well as their habitability. Many approaches have been adopted to determine the stability of these systems, including brute-force N-body simulations, semi-analytical calculations, and more recently machine learning methods. This allows significant advances in our understanding of planetary system dynamics, as well as providing tools to constrain unknown parameters of exoplanetary systems (assuming these systems are stable). In the following, we focus on planets around single star hosts, and we provide an overview of the studies of planetary system stability for compact multi-planet systems and hierarchical multi-planet systems. 
\end{abstract}

\begin{tcolorbox}
\textbf{Key Points} \\
\term{Reduced mass} It is defined as $m_1 m_2/(m_1 + m_2)$, where $m_1$ and $m_2$ are the masses of two objects. This allows the two-body problem to be treated as an equivalent one-body problem with the reduced mass moving in the potential field created by the two bodies.\\
\term{Separatricies} Boundaries in a dynamical system that separate regions of qualitatively different behavior in the phase space. \\ 
\term{Precession} The gradual shift or rotation of the orientation of an orbiting body's elliptical path around its primary body. \\
\term{Evection resonance} A gravitational interaction that occurs when the orbital precession rate of a satellite, such as the Moon, matches the average angular motion of its primary body, such as the Earth, around the Sun. This resonance can lead to significant changes in the satellite's eccentricity over time.\\
\term{Eviction resonance} Similar to Evection resonance, while the dominant effect is on inclination. \\
\term{Quadrupole order} Truncation at the second order in the ratio of the inner and outer binary semi-major axes ($a_{in}/a_{out}$) for hierarchical triple systems.\\
\term{Octupole order}  Truncation at the third order in the ratio of the inner and outer binary semi-major axes ($a_{in}/a_{out}$) for hierarchical triple systems.\\
\term{Hexadecapole order} Truncation at the fourth order in the ratio of the inner and outer binary semi-major axes ($a_{in}/a_{out}$) for hierarchical triple systems.\\
\term{Mean motion resonances} MMRs are integer ratios between the orbital periods of two planets. \\
\term{Hill radius} Characteristic radius around planets within which their own gravity dominates relative to the gravity of the central star. \\
\term{Secular resonances} Resonances that correspond to precession of the orbital planes of the planets.\\
\term{Angular Momentum Deficit} The difference between the angular momentum of an idealized system, in which the same planets of a real system are orbiting at their respective semi-major axes on circular and planar orbits, and the angular momentum of the real system. \\
\end{tcolorbox}

\section{Introduction}

The stability of planetary systems has been a subject of interest since the era of Johannes Kepler (1571-1630), who initially hypothesized that the spacing of planets in the Solar System was governed by the packing of perfect solids. Though this idea was later discarded, the fundamental questions about the packing and stability of planetary systems persist. Specifically, understanding how tightly planets can be packed in a system and the long-term implications of their gravitational interactions remains crucial. Small perturbations by nearby planets could either average out over long timescales, maintaining stability, or accumulate constructively, leading to orbital crossing and system instability through planetary collisions or ejections. It is thus a key question to determine which planetary systems are stable, and which are unstable (i.e., leading to scattering, collision and ejection of planets in the system). 

Significant progress has been made over the past decades, largely motivated by the discovery of thousands of exoplanetary systems. Many of the observed multi-planetary systems are compact, where planet-planet interactions are strong and instability is common. In this case, stability is a key factor in determining the properties of these systems. On the other hand, planetary systems with large spacing between the planets are generally stable, however, high mutual inclination between the planets may lead to secular resonances that causes high eccentricity excitation, which may then render the planet orbit unstable. Various approaches have been developed to analyze planetary system stability, ranging from analytical perturbative methods and brute-force N-body numerical simulations to novel machine learning techniques, facilitated by recent advancements in computational capabilities. In this review, we address these advances in our understanding of planetary system stability. We focus on compact multi-planetary systems in section 2 and hierarchical systems in section 3.

\section{Stability in compact multi-systems}\label{chap1:sec1}

\subsection{Packing of planetary systems and effects of resonances}


Spacing between planets plays a key role in the stability of planetary systems, and thus one useful tool used to help predict stability of a system is the mutual Hill radii between planets. The Hill radius is the characteristic radius around planets which their own gravity dominates relative to the gravity of the central star. The mutual Hill radii of two planets is defined as:
\begin{equation}
    R_H=\frac{a_1+a_{2}}{2}\left(\frac{m_1+m_{2}}{3M_*}\right)^{1/3}
\end{equation}
where $a_1$ is the semi-major axis of the inner planet, $a_{2}$ is the semi-major axis of the outer planet, $m_1$ and $m_{2}$ are the masses of the inner and outer planets respectively, and $M_*$ is the mass of the central star. In this unit, planetary separation can be expressed as a dimensionless number $K$:
\begin{equation}
    K=\frac{a_{2}-a_1}{R_H} .
\end{equation}


Although the stability of planetary systems is complex, analytical results do exist for simplified systems with circular and co-planar orbits. A well known result is that a particle with zero mass (otherwise known as a "test-particle") perturbed by a planet is subject to orbital chaos if it lies too close to the planet \citep{wisdom_resonance_1980}. Analytical studies of these systems show that overlap between mean motion resonances (MMRs) drive chaos in these systems which can lead to instability. Specifically, MMRs are integer ratios between the orbital periods of two planets, with inner orbital planet's period denoted as $P_1$ and outer orbital planet's period denoted as $P_2$. The dominant resonances in this regime are the first order MMRs when the planetary period ratio $P_2/P_1\sim(n+1)/n$, where $n$ is a positive integer. Based on overlapping of first order mean-motion resonances, \cite{wisdom_resonance_1980} derived a critical separation of
\begin{equation}
\frac{|a_{2}-a_{1}|}{(a_{2}+a_{1})/2}\leq1.49\mu^{2/7} ,
\label{sep_critic}
\end{equation}
where $\mu \equiv (m_{1}+m_{2})/{M_*}$ is the planet mass ratio, and the numerical coefficient is obtained from \citep{Duncan89}. In K value, the critical spacing is 
\begin{equation}
K_{crit} = 3.46\Big(\frac{\mu}{4.5\times10^{-5}}\Big)^{-1/21} .
\end{equation}
Relaxing the assumption of a test particle, \cite{deck_first-order_2013} considered planetary systems with two massive planets on nearly circular and nearly coplanar orbits near a first-order resonance. The Hamiltonian governing the dynamics can be reduced to a one-degree-of-freedom system when the orbital eccentricities and inclinations are small. The critical spacing based on resonance overlap criterion is little modified if the test particle is replaced by a planet.

\begin{figure}
    \centering
    \includegraphics[width=0.95\textwidth]{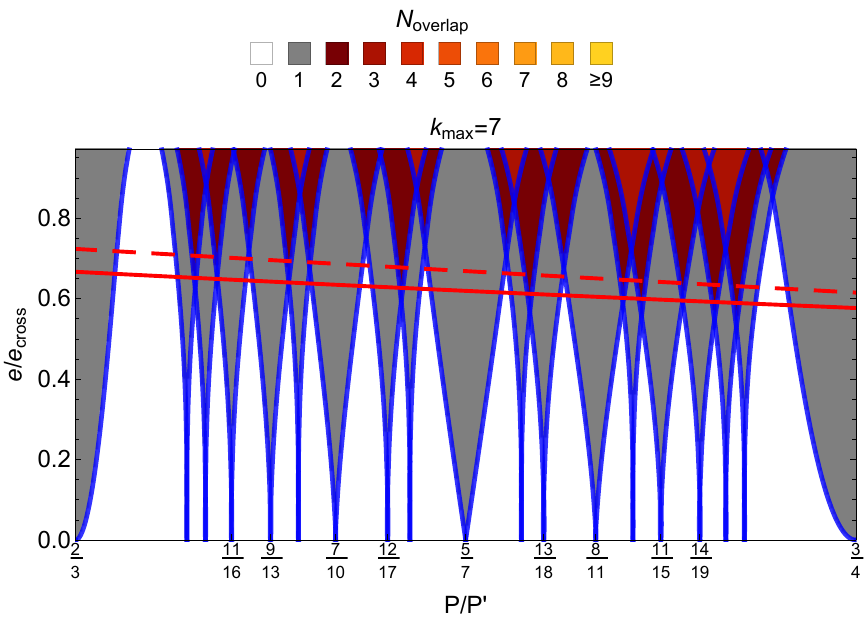}
    \caption{Structure of resonances and their overlap between the $P_2/P_1=3/2$ and $P_2/P_1=4/3$ MMRs for an inner test particle perturbed by an outer planet on a circular orbit. The mass of the outer planet is taken to $10^{-5}$ times the mass of the central star. Blue separatricies are plotted for resonances up to seventh order. The x-axis shows the ratio of periods of the inner test particle $P$ to the outer planet $P'$. The y-axis shows the eccentricity of the inner test particle in terms of $e_{cross}$ which is the eccentricity at which the orbits cross. $e_{cross}\approx0.25$ at these periods. The color corresponds to the number of resonances which overlap at the location. The solid red line corresponds to the approximate eccentricity at which the number of overlapping resonances is one. The dashed red line marks the estimate from \cite{hadden_criterion_2018} where chaos and instability occur. Figure from \cite{hadden_criterion_2018}.}
    \label{fig:mmr_overlap}
\end{figure}

More recently, this result has been generalized for eccentric orbits \citep{hadden_criterion_2018}. Figure \ref{fig:mmr_overlap} shows an example of resonances overlapping in the two-planet case where the inner planet has a non-zero eccentricity. For this illustrative example the mass of the inner planet is assumed to be a test particle and the mass of the outer planet is taken to $10^{-5}$ times the mass of the central star. Specifically, what is shown are the locations and widths for all resonances up to $7$th order between the $P_2/P_1=3/2$ and $P_2/P_1=4/3$ MMRs at different inner planet eccentricities. At low inner planet eccentricities, the resonances are narrow and there is no overlap. As inner planet eccentricity increases, the resonances widen and overlap exists everywhere. For any given $P_2/P_1$, there is a critical eccentricity above which the inner test particle comes under the influence of two resonances simultaneously, leading to chaotic dynamics.

Adding just a single planet to this simplified model greatly increases the complexity of the problem and instability of the system. For systems with three or more planets, instabilities can occur even when the separations between adjacent planet pairs far exceed the two-planet limit given in Eq. (\ref{sep_critic}). More planets means a larger number of available MMRs between all possible pairs of planets. In addition, three-body resonances (integer ratios between the periods of three planets) act to ``fill the space'' between two-body MMRs. This intricate overlapping of resonances drives instabilities over a continuous range of separations \citep[e.g.,][]{quillen_three-body_2011}. Recently, \cite{Rath22} showed that predictions made considering the two-body (including those not overlapping) and three-body resonances allow detailed agreement with high-resolution maps of chaos from N-body simulations. Generalizing to systems with a large number of planets and applying to the observed Kepler systems, direct N-body simulations have been adopted to characterize the stability limit. This gives $K_{crit} \sim 10-12$ for a stable timescale of $Gyr$, agreeable with observations (see more discussions in \ref{sec:inst}). 

In addition to MMRs, there exist yet more resonances which may lead to instability within a system. If the effects of MMRs are removed, then planets will exchange angular momenta between each other's orbits at fixed semi-major axes \citep{murray_solar_1999} and resonances may develop between the slower set of frequencies that correspond with the precession of the orbital planes of the planets \citep{lithwick_theory_2011}. Overlap of these so-called secular resonances can drive chaos in the system and also lead to instabilities. However, within these secular dynamics, there exists a conserved quantity called the Angular Momentum Deficit (AMD) which can be used to set some limits on the instability of a system. The AMD is defined as the difference between the angular momentum of an idealized system, in which the same planets of a real system are orbiting at their respective semi-major axes on circular and planar orbits, and the angular momentum of the real system. In other words, AMD is the projection of the orbital angular momentum vector of the real system onto the invariable plane of that system \citep{laskar_chaotic_1990}. As planets exchange angular momentum, through the above mentioned secular resonances, the AMD acts as a constant reservoir of eccentricity and inclination that the planets can trade with one another. The AMD of a system can therefore act as a test for the secular instability of a system. For example, if the AMD is too small to allow for orbit crossing and planet collisions when all of the available eccentricity is given to one adjacent pair of planets, then the system is called AMD stable.

Though using the AMD of a system to asses stability can be a simple and powerful analytic tool, it has two important limitations. First, it provides no information on the instability time of a system that is classified as AMD unstable. This is due to the fact that AMD stability is a classification for stability. A famous example is our own Solar System which is classified as AMD unstable. However, most $(\approx99\%)$ direct simulations of our Solar System predict it will remain stable over the main sequence lifetime of the Sun.
Second, it provides constraints only on the secular dynamics of a system and necessarily ignores the effects of MMRs. As mentioned before, MMRs are an important source of dynamical chaos for compact systems. 

\subsection{Instability timescale and indicators}
\label{sec:inst}
How long do planetary systems stay stable? Several studies have measured instability times of systems with different planetary separations using large suites of direct N-body integrations. They find that instability times rise steeply with increasing separation between planets measured in mutual Hill radii\footnote{Some researchers \citep[e.g.,][]{quillen_three-body_2011} 
argue that instability times better scale with the planet-star mass ratio $\mu$ to the one fourth power, rather than the traditional $\mu^{1/3}$ Hill scaling that we have shown here. However, as \cite{tamayo_predicting_2020} points out, these scalings are close together and using Hill radii to determine instability times is inaccurate compared to more sophisticated numerical models.} \citep[e.g.,][]{chambers_stability_1996}
In short and in general, the more widely planets are spaced, the longer the system remains stable. However, other orbital parameters besides interplanetary spacing can also strongly influence system stability. Several studies have also considered the effects of finite planet eccentricities and inclinations in addition to interplanetary spacing to predict instability times \citep[e.g.,][]{yoshinaga_stability_1999}
. These studies usually make various simplifying assumptions in their work such as equally spaced planets, equal planetary eccentricities, etc. These different assumptions lead to different reported quantitative relationships between planet spacing and instability times and so the exact correlation remains unclear. In addition, real observed exosystems are rarely as uniform as these simplified models and so the applicability of their predictions to real systems remains in question.

The most straightforward way to predict the instability time of a given system is by direct numerical integration of the equations of motion of the system and simulating its evolution.\footnote{Even this straightforward method is not perfect. It has been shown than an N-body integration can be interpreted as sampling a single instability time from a broader distribution of values.} 
Many researchers have used this technique to narrow down physical orbital architectures and formation histories for important exoplanet discoveries. They accomplish this by discarding configurations that lead to rapid instability in their simulations since one does not expect to discover a system just prior to chaotic collapse \citep[e.g.,][]{steffen_transit_2013}.
However, this method tends to be computationally expensive, especially for exo-systems with planets whose orbital parameters are poorly constrained. Due to the high dimensionality of the parameter space, usually only a small fraction of proposed orbital configurations for these systems are explored and integration timespans are typically many orders of magnitude shorter than the expected $\sim$ Gyr lifetimes of such systems.

Computation time scales linearly with the number of orbits and requires $\sim10$ s per million orbits with optimized algorithms \citep{Wisdom1991} and modern hardware \citep{tamayo_predicting_2020}. Luckily, for some exo-systems this computational expense in within reach. For example, the system HR 8799 has an innermost planet with an orbital period of $\approx40$ yrs and requires only $10^6$ orbits to simulate the $\sim40$ Myr age of the system \citep{wang_dynamical_2018}. Unfortunately, young multi-planet systems with long inner orbital periods like HR 8799 are quite rare in the current catalogue of discovered exoplanets. A combination of population statistics and strong observational biases results in the discovery of multi-planet systems which most frequently have innermost orbital periods of $\approx0.01 - 0.1$ yrs. Many of these systems commonly surround stars that are several Gyrs old and so it is typically necessary to simulate $10^{11} - 10^{12}$ orbits to fully study the stability of such systems. In these cases it becomes prohibitively computationally expensive to test all proposed orbital configurations which are typically necessary to evaluate stability.  

Researchers have thus been motivated to find alternative numerical methods that identify system instability with less computational expense. One popular way is to use chaos indicators that are numerically measured from short integrations of a system. These indicators provide some insight on how quickly chaos will develop within a system for a given orbital configuration \citep[e.g.,][]{marzari_dynamical_2014}. A common chaos indicator used for studying compact planetary systems is the Mean Exponential Growth factor of Neaby Orbits (MEGNO) \citep{cincotta_phase_2003}. This approach is useful since unstable systems will usually display chaotic characteristics on short timescales and can be identified with little computational expense. However, this method is not perfect. It is possible for a planetary system to be chaotic yet never become unstable on astrophysically relevant timescales. Therefore, chaos indicators like MEGNO may incorrectly infer a system's instability based on the mere existence of chaotic dynamics within the system. In addition, measurement of chaos indicators based upon short integrations will fail to measure chaos on timescales longer than those simulated. This means that measuring chaos indicators on short timescales may misclassify systems that are destabilized as a result of chaotic dynamics with longer timescales.

\subsection{Advances using Machine Learning}
Machine Learning (ML) provides an alternative method and has allowed researchers to numerically classify the stability of solar systems with significantly lower computational cost. For exmaple, \cite{tamayo_predicting_2020} developed a machine learning model called SPOCK (Stability of Planetary Orbital Configurations Klassifier) that can reliably classify the stability of compact solar system configurations with three or more planets over $10^9$ orbits. This is two orders of magnitude less than the aforementioned target of $10^{11}$ orbits usually needed to fully rule out instability. However, when researchers discover an exosystem with a poorly constrained configuration, this tool allows them to quickly reject configurations that lead to instabilities on short timescales. Even if a discovered system has a dynamical age of $>10^{11}$ orbits, this technique can greatly reduce the parameter space of proposed configurations given the low likelihood of finding a system within $<1\%$ of its lifetime of going unstable. One of the main advantages of using the machine learning model SPOCK is that it can be up to $10^5$ times faster to evaluate the stability of a particular configuration compared to directly simulating that same system through numerical integration.

\begin{figure}
    \centering
    \includegraphics[width=0.95\textwidth]{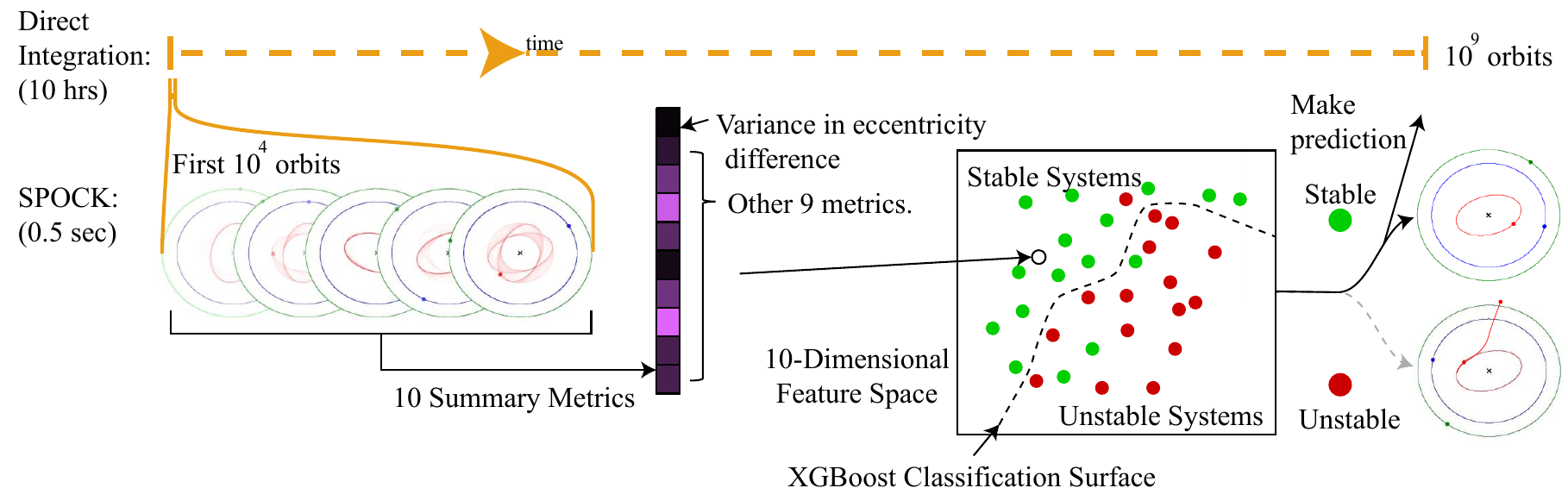}
    \caption{An illustration showing how ML models like SPOCK can classify the stability of an initial system configuration. The traditional approach numerically integrates the system for $10^9$ orbits, requiring up to 10 hours with modern hardware. SPOCK runs a much shorter $10^4$ orbit integration, and from it generates a set of 10 summary features. These map to a point (white circle) in a 10-dimensional space, in which SPOCK has been trained to classify stability using a machine learning model called XGBoost. SPOCK outputs an estimated probability that the given system is stable over $10^9$ orbits, up to $10^5$ times faster than direct integration. Figure modified based on Figure 1 from \cite{tamayo_predicting_2020}.}
    \label{fig:spock_cartoon}
\end{figure}

ML models like SPOCK are trained on large datasets which are generated from the results of computationally expensive direct numerical integrations. These datasets are created from a large suite ($\gtrsim10^5$) of initial conditions for solar system configurations that are simulated for $\sim10^9$ orbits. The input for these models are the complete initial orbital configuration of the system, which includes the masses of the central star and planets as well as the positions and velocities for each planet (or equivalently their orbital elements). A relatively short and inexpensive integration is performed ($10^4$ orbits) and dynamically informative quantities are measured from these results. Based upon the measurements of these quantities, the model evaluates the stability of the system. Figure \ref{fig:spock_cartoon} shows an illustration of this process.

\begin{figure}
    \centering
    \includegraphics[width=0.5\textwidth]{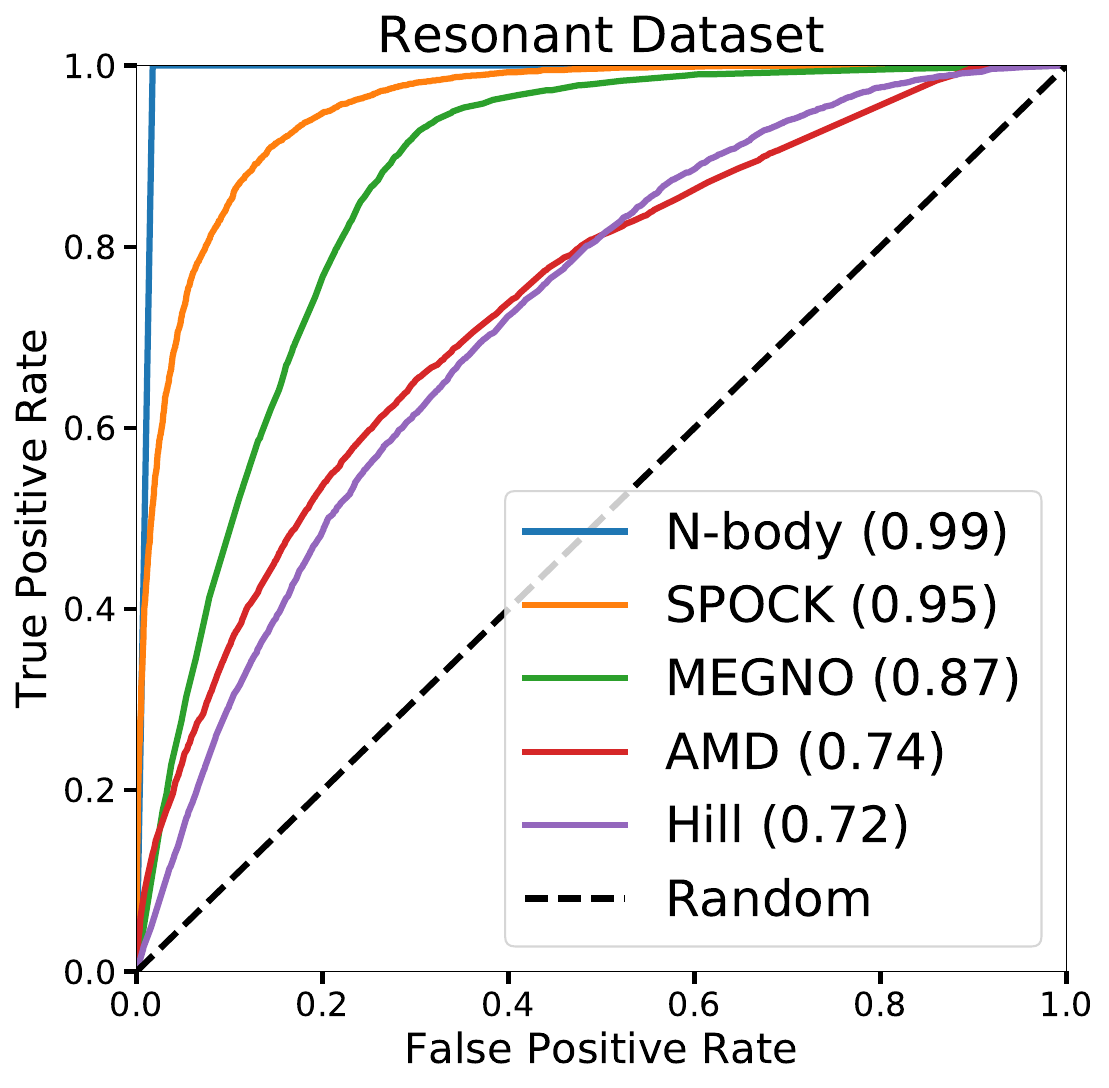}
    \caption{Comparison of the performance of SPOCK against previous models. Plots True Positive Rate (TPR: fraction of stable systems correctly classified) vs. False Positive Rate (FPR: fraction of unstable systems misclassified as stable). All models can trade off TPR vs. FPR by adjusting their threshold for how conservative to be before labeling a system as stable. The area under the curve (AUC) for each model is listed in the legend in parentheses. A perfect model would have a value of unity, and random guessing (dashed black line) would have AUC=0.5. The blue N-body curve gives an empirical estimate of the best achievable performance on this dataset. At an FPR of 10\%, SPOCK correctly classifies 85\% of stable systems, MEGNO 49\%, AMD 36\%, and Hill 30\%. Figure from \cite{tamayo_predicting_2020}.}
    \label{fig:spock_perform}
\end{figure}

The specific quantities that are measured by ML models to evaluate a system's stability depends on the specific model being used. The way these quantities are assessed also differs by model. For example, \cite{tamayo_predicting_2020} use features selected by the authors informed by analytical insights. These features include the above mentioned MEGNO chaos indicators as well as the locations and strengths of the nearest MMRs for each adjacent planet pair. The selection of each feature and their relative assigned importance strongly correlate to the performance of the ML model. The assigned importance of each feature is adjusted as the model iteratively trains on its training set until accuracy is optimized. Figure \ref{fig:spock_perform} shows the performance of SPOCK compared to previously discussed numerical models when binarily classifying a given system configuration as "stable" or "unstable" over $10^9$ orbits. While SPOCK is less accurate than direct numerical integration, it and similar ML models greatly outperform previous models for significantly less computational expense than fully simulating a system's lifetime.

Instead of relying upon scientist-derived instability metrics, more recent ML models are able to learn their own metrics from scratch. This approach not only improves on long-term predictions of previous models based on engineered features, but also significantly reduces the model bias and improves generalization to types of solar system configurations not covered in the model's training set. For example, \cite{cranmer_bayesian_2021} designed a Bayesian neural network which incorporates these ideas and is able to estimate the instability time of a given system configuration. The instability times predicted from this model are up to two orders of magnitude more accurate at predicting instability times than analytical estimators.

\section{Stability in hierarchical systems}\label{chap1:sec2}

\subsection{Dynamics of hierarchical triple systems}

\begin{figure}
    \centering
    \includegraphics[scale=0.5]{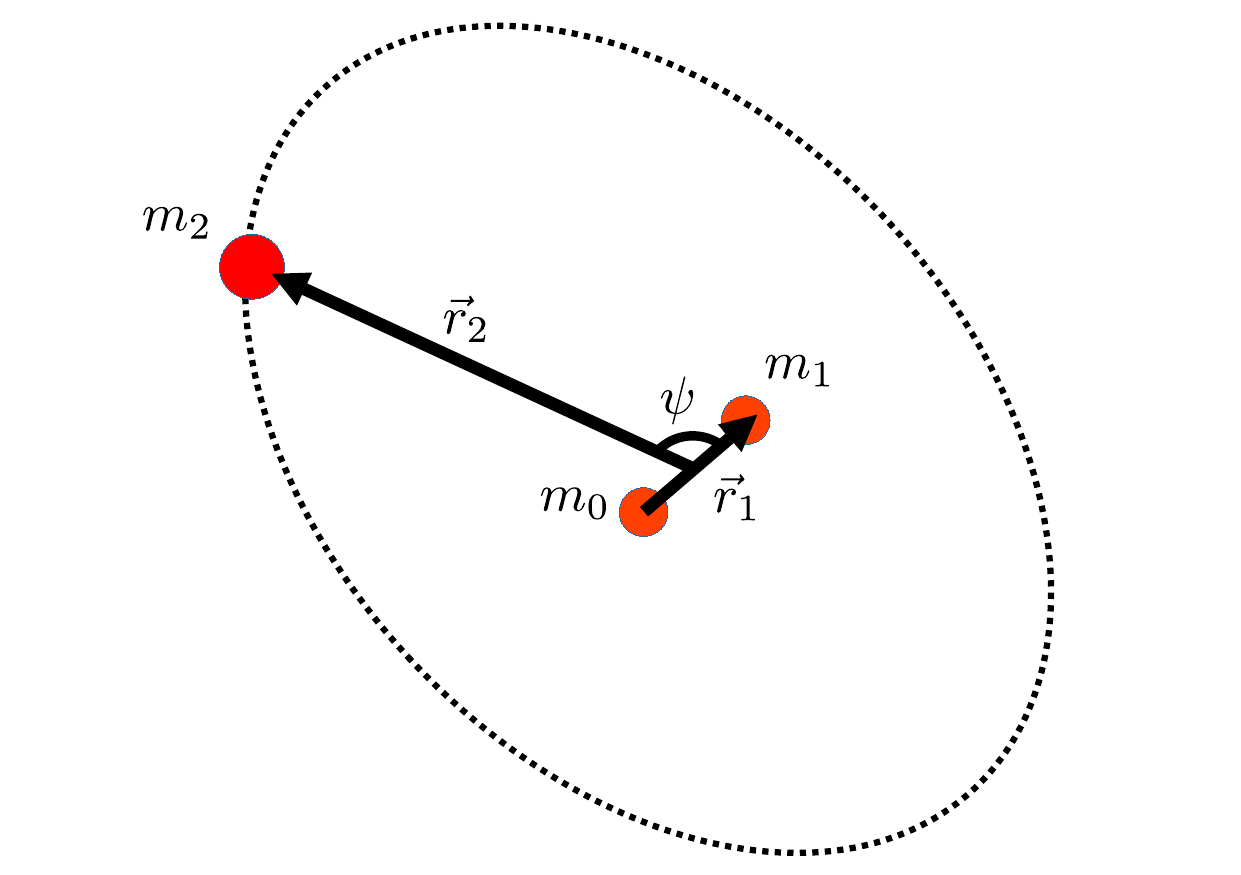}
    \caption{Schematic of a hierarchical three body system. Masses $m_0$ and $m_1$ form the inner binary and $m_2$ is the distant companion. $\vec{r}_1$ is the relative position vector of $m_1$ with respect to $m_0$. $\vec{r}_2$ is the relative position of $m_2$ with respect to the center of mass of the inner binary, and $r_1\ll r_2$. $\psi$ is the angle between $\vec{r}_1$ and $\vec{r}_2$.}
    \label{fig:schHierar}
\end{figure}

Hierarchical systems (nested systems where the separations between bodies are widely different, see the configuration in Figure \ref{fig:schHierar}) are ubiquitous in nature; many astrophysical systems including planetary and stellar systems can be modeled as hierarchical systems. The most abundant and well studied of hierarchical systems are hierarchical triple systems. A hierarchical triple system consists of an inner binary and a perturber orbiting the center of mass of the inner binary, where the perturber is on a much wider orbit as compared to the orbit of the inner binary \citep{eggletonEmpiricalConditionStability1995a}.

We now define parameters of a hierarchical three body system (see Figure \ref{fig:schHierar}). Masses $m_0$ and $m_1$ form the inner binary. The mass $m_2$ is orbiting the center of mass of the inner binary on a much wider orbit. The orbital elements of the inner (outer) orbit are given by: semi-major axis $a_1$ ($a_2$), eccentricity $e_1$ ($e_2$), inclination $i_1$ ($i_2$), argument of pericenter $\omega_1$ ($\omega_2$), longitude of ascending node $\Omega_1$ ($\Omega_2$) and mean longitude $\lambda_1$ ($\lambda_2$). The relative position vector of mass $m_1$ with respect to $m_0$ is given by $\vec{r}_1$, and the relative position of $m_2$ with respect to the center of mass of $m_0$ and $m_1$ is given by $\vec{r}_2$. The mean motion of the inner (outer) orbit is $n_1 = \sqrt{G(m_0+m_1)/a^3_1}$  ($n_2 = \sqrt{G(m_0+m_1+m_2)/a^3_2}$). The system is considered hierarchical if the semi-major ratio $\alpha = a_1/a_2 << 1$. It should be noted that despite the three subscript labels, there are only two osculating Keplerian orbits, namely the inner and the outer orbits. In practice, techniques used to study hierarchical systems are valid for $\alpha \leq 0.1$ \citep{naozEccentricKozaiLidovEffect2016}.


The types of interactions in a three body system can be broadly classified into three categories: close encounters, resonant interactions including MMRs, and secular interactions (e.g., \cite{murray_solar_1999}). Close encounters occur on dynamical timescales ($ t_{p,1}= 2\pi/n_1$), and can quickly destabilize a triple system. MMRs are important when the periods of the inner and the outer orbits are commensurable. The nominal location of a $p:q$ MMR is given by the relation:  $n_1/n_2 = p/q \implies a_2 \sim a_1 \left( \frac{p}{q} \right)^{2/3} $. MMRs operate on timescales roughly given by $t_{res} \sim \sqrt{m_2/(m_0+m_1)} t_p$ \citep{murray_solar_1999}. Resonant interactions are important only within a certain distance (called resonance widths) from their nominal locations. It can be shown that the resonance width of a $p:q$ MMR reduces with $p-q$ (also called the order of the resonance). Consequently, it can be shown that MMRs are mainly important for $\alpha >= 0.1$. Since the inner and the outer binaries are widely separated in hierarchical systems ($\alpha < 0.1$), MMRs do not play a significant role in their dynamics. Secular perturbations on the other hand can significantly influence the long term evolution of a hierarchical triple system (e.g., \cite{naozSecularDynamicsHierarchical2013}). In general, secular perturbations operate on much longer timescales given by: $t_{sec} \sim (m_0/m_2)(a_2/a_1)^3 t_p$. It should be noted that under certain conditions resonant interactions like evection resonances can also be important in hierarchical triples (e.g., \cite{toumaResonancesEarlyEvolution1998}). As we will see later, long term dynamics dictated by secular or resonant interactions can lead to close encounters, which can then destabilize a hierarchical triple system.

Perturbation theory can be used to study the secular and the resonant dynamics of triple systems. In a hierarchical system, the perturbing potential ($\Phi$) of the third body can be expanded in the ratio of the semi-major axis of the inner and the outer orbits ($\alpha$) (e.g., \cite{harringtonStellarThreeBodyProblem1969}). The perturbing potential can be written as:
\begin{equation}
    \Phi_i = \frac{G m_j}{|\vec{r}_2-\vec{r}_1|} = \frac{G m_j}{|\vec{r}_2|} \sum_{k=1}^{\infty} P_k(\cos{\psi}) \left(\frac{|\vec{r}_1|}{|\vec{r}_2|}\right)^k
\end{equation}
where $i,j \in \{1,2\}, i \neq j$, ${\psi} = \arccos({\hat{r}_1.\hat{r}_2})$ is the angle between the relative position vectors $\vec{r}_1$ and $\vec{r}_2$, and $P_k (x)$ is the Legendre polynomial of degree $k$. 
To study secular dynamics, the system can be further simplified by averaging  $\Phi$ over the mean longitude of the inner and the outer orbits. Averaging over the mean anomalies of the orbits removes short-term interactions between the planets, and helps to focus on the long term dynamics of the two planets. In addition, it can be shown that double averaging is equivalent to performing a canonical transformation which eliminates the mean anomalies of the two orbits from the Hamiltonian of the three body system (e.g., \cite{naozSecularDynamicsHierarchical2013}).
\begin{equation}
    \Phi_{da,i} = \frac{1}{4\pi^2} \int_0^{2\pi} \int_0^{2\pi} \Phi_{i} d\lambda_1 d\lambda_2
 \end{equation}
It should be noted that the secular double averaged approximation treats the inner and the outer orbits as massive interacting rings with constant semi-major axis. Also, double averaging eliminates all resonant terms from the perturbing potential. 


In a hierarchical triple system with the inner binary containing a test particle, the dynamics is further simplified if the eccentricity of the outer orbit is taken to be zero. If the secular approximation is valid, an expansion of the double averaged perturbing potential up to quadrupole order in the ratio of semi-major axis can be used. A closed form analytical solution to the secular equations of motion can be derived to obtain explicit functions of $e_1(t)$ and $i_1(t)$ 
. The resulting dynamics is described by the von Zeipel-Lidov-Kozai (vZLK) Mechanism \citep{vonZeipel1910,Lidov1962,Kozai1962}. In such systems, if the mutual inclination between the initially near circular inner and the outer orbits ($i_m$) is greater than $\arccos(\sqrt{3/5}) \sim 39.2^\circ$, the eccentricity of the test particle can be periodically excited to large values on secular timescales. The system also has a constant of motion: the component of orbital angular momentum of the test particle along the orbit normal of the outer orbit ($h=\sqrt{1-e_1^2} \cos{i_m}$).  Starting from a near circular orbit, the eccentricity of the test particle can be excited to a maximum of $e_{max} = \sqrt{1-\frac{5\cos^2{i_{m,0}}}{3}}$, where $i_{m,0}$ is the mutual inclination when the eccentricity of the inner orbit is near zero.

\begin{figure}
    \centering
    \includegraphics[scale=0.8]{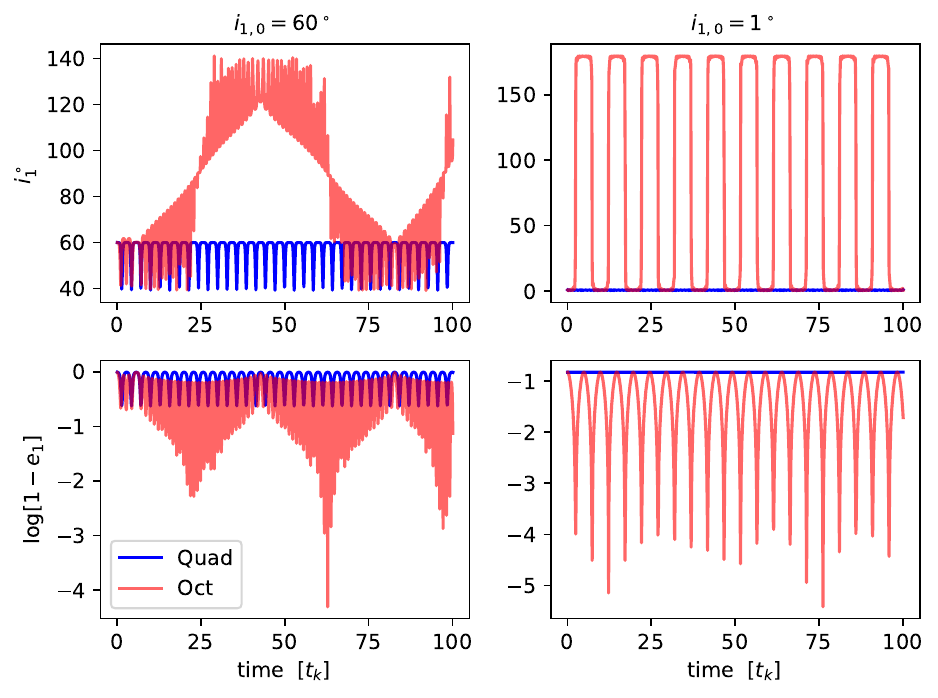}
    \caption{The evolution of the orbit of a test particle as calculated by quadrupole (shown in blue) and octupole (shown in red) expansions of the perturbing potential. Here, $t_{k} = (m_0/m_2)(a_2/a_1)^3 t_{p,1}$. Comparing the red and the blue trajectories we can see that the quadrupole order expansion cannot be used for these initial conditions. We use the following initial conditions to make this figure: Left panel: $\alpha = 0.022$, $e_1=0.01$, $i_1=60^\circ$, $\omega_1=0$ and $\Omega_1=60^\circ$, Right panel: $\alpha = 0.022$, $e_1=0.85$, $i_1=1^\circ$, $\omega_1=0$ and $\Omega_1=180^\circ$. For the blue (red) trajectories $e_2=0$ $(e_2=0.85)$. Adapted from \cite{naozEccentricKozaiLidovEffect2016} with modified initial conditions.}
    \label{fig:trajOctQuad}
\end{figure}

Figure \ref{fig:trajOctQuad} shows the orbital evolution of a test particle orbits perturbed by a distant companion. The top (bottom) rows show the evolution of the inclination (eccentricity) of the test particle. The trajectories in blue show the evolution as calculated by the quadrupole order expansion of the perturbing potential. The initial eccentricities are set close to zero. The initial mutual inclination in the left (right) column is 60$^\circ$ (1$^\circ$). We can see that the eccentricity is excited to 0.76 in the left panel. Meanwhile, the inclination oscillates between 60$^\circ$ and 40$^\circ$. The eccentricity and the inclination of the test particle remain close to their initial values, and are not excited in the right panel.

The original vZLK mechanism, which is applicable to binary systems with a circular outer companion, has been extended to study more general configurations of hierarchical triple systems. For instance, eccentric vZLK  mechanism describes the secular evolution of a binary system with an eccentric outer companion \citep{naozEccentricKozaiLidovEffect2016}. The perturbing potential must be expanded up to at least octupole order in the semi-major axis ratio ($\alpha^3$) to study the secular evolution induced by an eccentric companion. The octupole order term is proportional to $\epsilon=\alpha e_2/(1-e_2^2)$, and the octupole order expansion is valid up to $\epsilon<0.1$. Unlike the circular companion case, the system does not have a constant of motion other than the Hamiltonian. The system now has two degree of freedom system with no closed form analytical solution. Consequently, numerical integrations have been used to study their dynamics. It has been shown that with an eccentric companion, the eccentricity of the test particle can be excited to values close to 1 for a wider range of initial inclinations. In addition, the orbit of the test particle can flip from prograde to retrograde and vice versa \citep{liChaosTestParticle2014,liEccentricityGrowthOrbit2014}. Figure \ref{fig:trajOctQuad} also shows the octupole order evolution of a test particle orbit (in red). For the initial conditions chosen in the left panels, we can see that the eccentricity is excited to values close to 1. Also, the orbit of the test particle flips from prograde ($i_{m}<90^\circ$) to retrograde ($i_{m}>90^{\circ}$). In the right panels, we can see that in the octupole approximation, the eccentricity can be excited even for initially near coplanar orbits ($i_{m} \sim 0^\circ$). The eccentricity excitation is accompanied by the flipping of the orbit from prograde ($i_{m} \sim 0^\circ$) to retrograde ($i_{m} \sim 180^\circ$). Compared to the quadrupole order evolution (shown in blue), we can see that the eccentricity can be excited to larger values for systems with eccentric companions.

In our discussion so far we have focused on the Newtonian dynamics of three point mass objects. This may be inadequate for highly eccentric orbits. More specifically, eccentricity excitation caused by secular three body interactions allows close pericenter passage where additional effects like general relativity, tides and rotational deformation can be important. The additional precession induced by these short range forces suppresses the eccentricity excitation. While the mutual inclination window where eccentricity excitation is possible is not significantly affected by short range forces, the parameter space where orbital flips are possible is restricted \citep{liuSuppressionExtremeOrbital2015}.  

Higher order expansions including up to hexadecapole order of the perturbing potential have also been derived in literature (e.g.,\cite{willOrbitalFlipsHierarchical2017,vinsonSecularDynamicsExterior2018}). Such expansions can be used to study less hierarchical systems.  At higher values of $\alpha$, the dynamics tends to be more chaotic \citep{vinsonSecularDynamicsExterior2018}. A larger number of secular resonances become important at higher values of $\alpha$, and the overlap of these secular resonances leads to more chaotic regions. A few studies have also explored the dynamics of non-hierarchical triple systems. Instead of expansion in $\alpha$, they numerically average the perturbing potential (e.g., \cite{beustOrbitalClusteringDistant2016,bhaskarMildlyHierarchicalTriple2021}). This approach is valid for any value of $\alpha$, except where MMRs or close encounters are important.

The secular approximation is only valid for weak pertubations. Under certain conditions, when the companion is massive, the secular approximation can underestimate the maximum eccentricity attained by the binary. A more accurate calculation involves resolving the orbit of the companion. It has been shown that a single averaged perturber potential can be used, instead of the secular double averaged potential, to properly account for the long term evolution of the binary \citep{luoDoubleaveragingCanFail2016,grishinQuasisecularEvolutionMildly2018}. Generally, the secular approximation is valid as long as the ratio of the mass of the perturber and the mass of the central object is less than $10^{-3}$ for $\alpha<0.3$ \citep{bhaskarMildlyHierarchicalTriple2021}.

\subsection{Stability of hierarchical triple systems}
Multiple studies have analyzed the stability of Hierarchical triple systems \citep{georgakarakosStabilityCriteriaHierarchical2008}. Both analytical and empirical stability criteria have been derived. In this section, we will mainly focus on empirical stability criteria for Hierarchical triple systems. Most algebraic stability criteria derived in literature can written in the following form:
\begin{equation}
    r_{stab} = \frac{a_{2}(1-{e_{2}})}{a_{1}(1+e_{1})} > Y
\end{equation}
where $Y$ is a function of the masses and the initial orbital elements of the inner and the outer orbits. In addition, $a_1$ ($a_2$) and $e_1$ ($e_2$) are the initial semi-major axis and the eccentricity of the inner (outer) orbits.  We list the functional forms of $Y$ derived in literature in Table \ref{tab:stabcrit}. Here, $\mu_1=m_1/m_0$ and $\mu_1=m_2/m_0$ are the mass ratios and $i$ is the initial mutual inclination between the inner and the outer binaries. In the following we discuss the results in greater detail.

\begin{table}
    \centering
    
    \caption{List of empirical stability criteria derived in literature for planetary systems. Here, $a_1$ ($a_2$) is the initial semi-major axis, $e_1$ ($e_2$) is the initial eccentricity of the inner (outer) orbit, $\mu_1=m_1/m_0$ and $\mu_2=m_2/m_0$ are the mass ratios and $i$ is the initial mutual inclination between the inner and the outer binaries. The triple is stable if $r_{stab}=\frac{a_{2}(1-{e_{2}})}{a_{1}(1+e_{1})}  > Y$. }
    \begin{tabular}{|c|c|c|}
        \hline
        Citation & $Y$ & Applications\\
        \hline
         \cite{harringtonPlanetaryOrbitsBinary1977}&\begin{tabular}{@{}c@{}} \\ $\frac{A}{(1+e_1)}\left(1+B\log{\left(\frac{1+(\mu_2/(1+\mu_1))}{3/2}\right)}\right)+2$, where \\ $A=3.5, B=0.7$ for $i=0$ and \\ $A=2.75, B=0.64$ for $i=180^\circ$ \\ \\ \end{tabular} &planetary systems \\
        \hline
         \cite{mardlingTidalInteractionsStar2001a} & \begin{tabular}{@{}c@{}} \\  $\frac{2.8}{1+e_{1}}\left( \left(1+\frac{\mu_{2}}{1+\mu_1}\right) \left( \frac{1+e_{2}}{(1-e_{2})^{1/2}} \right)\right)^{2/5} $ \\ \\ \end{tabular} & moderate mass ratios\\
         \hline
         \cite{grishinGeneralizedHillstabilityCriteria2017a} & \begin{tabular}{@{}c@{}} \\ $\frac{g(i)^{2/3}(1+e_{max}) \mu_2^{1/3}}{(1+\mu_1)^{1/3}}$, where \\ $g(i)=\cos{i}+(3+\cos^2{i})^{1/2}$ and \\ $e_{max}=\sqrt{1+(5\cos^2{i}/3)}$. Also, $e_2=0$. \\ (only vZLK correction)  \\ \\\end{tabular}& \begin{tabular}{@{}c@{}}Extreme mass ratios \\ $m_2>>m_0>>m_1$ \end{tabular}\\
         \hline
         \cite{toryEmpiricalStabilityBoundary2022} & \begin{tabular}{@{}c@{}} \\ $((1+e_{1})f(q)g(i)h(q,i))^{-1}$, where \\ $q=\frac{1+\mu_1}{\mu_2}$, $f(q) = 10^{-0.6+0.04q}q^{0.32+0.1q}$, \\ $\log(h(q,i))= -i q^{1.3}/1500$, \\ $g(i)=-0.4\cos{i} +1.4$ for $i<60^\circ$ and \\ $g(i)=-0.1773i^4 + 1.1211i^3-1.9149i^2+0.5022i+1.6222$ for $i>60^\circ$ \\ \\ \end{tabular}& Wide range of mass ratios\\
         \hline
         \cite{petrovichStabilityFatesHierarchical2015a} & \begin{tabular}{@{}c@{}} \\ $2.1\mu^{2/7}_{1}(a_{2}/a_{1})^{1/2}+1.03$ \\ \\ \end{tabular} & planetary systems\\
         \hline
         \cite{vynatheyaAlgebraicMachineLearning2022a} & \begin{tabular}{@{}c@{}} \\ Machine learning model \\ \\ \end{tabular} & moderate mass ratios\\
         \hline
    \end{tabular}
    \label{tab:stabcrit}
\end{table}
In an early study, \cite{harringtonStabilityCriteriaTriple1972} derive an empirical stability criteria using an ensemble of N-body simulations, with initial conditions sampled over a limited range of eccentricities and inclinations. The integrations were run for equal mass systems for 10-20 orbital periods of the inner binary. They define a triple system to be stable if there was no significant change in the orbital elements of the orbits. They found that the stability of the triple had the strongest dependence on the pericenter distance of the outer binary. The dependence on the inclination and other orbital elements was found to be weak. This is due the fact that N-body integrations used in this study were run for short timescales. Consequently, long term secular effects were not taken into account. Building on these results, 
\cite{harringtonPlanetaryOrbitsBinary1977} focus on the stability of a planet orbiting a single star (with a stellar companion), and a binary star system on coplanar prograde and retrograde orbits. A modified stability criteria was derived which had similar logarithmic dependence on the mass ratios of the inner and the outer binaries.


Some studies used a semi-empirical approach to derive stability criteria. For instance, \cite{mardlingTidalInteractionsStar2001a} drew analogy between the stability of hierarchical three body systems, and the stability against chaotic energy exchange in the binary-tides problem to derive a semi-empirical stability criteria. The criteria is applicable for systems on coplanar prograde orbits with arbitary eccentricities. They find that their formula is valid for mass ratio ($m_2/(m_0+m_1)$) less than 5, beyond which exchange between the inner and the outer binary is possible.

Focusing on satellites orbiting their host planets ($m_2 {\textgreater}{\textgreater} m_0 {\textgreater}{\textgreater}m_1$), \cite{innanenLimitingRadiiDirect1979a} derived a limiting Hill radius within which the satellite's orbit is stable. Using zero velocity curves of a circular restricted three body problem, the author shows that the limiting radius of retrograde orbits is larger than for prograde orbits. 
\cite{grishinGeneralizedHillstabilityCriteria2017a} then derived corrections to the stability limit by accounting for the eccentricity excitation caused by vZLK oscillations and Evection resonance. The corrected limiting radius is shown to agree much better with N-body integrations.

\cite{toryEmpiricalStabilityBoundary2022} focus on deriving an empirical stability criteria which is valid for arbitrary mass ratios and mutual inclinations. They attempt to bridge the extreme mass ratio regime, where the hill stability criteria is valid and equal mass regime, where stability limit is well modeled by \cite{mardlingTidalInteractionsStar2001a}. They find that the hill regime is valid for $(m_0+m_1)/m_2 <0.1$. Their fit predicts stability for $10^{4}$ orbits of the inner binary with an 88\% accuracy.

New computational techniques are being used to advance our understanding of the stability of hierarchical triple systems. More specifically, advances in machine learning algorithms have allowed researchers to learn interesting features of the non-linear multi-dimensional stability problem.  For instance, \cite{petrovichStabilityFatesHierarchical2015a} study the stability of hierarchical two planet systems using Support Vector Machine (SVM) trained on a large set of N-body simulations. Their N-body simulations run for a maximum of $10^8$ orbits of the inner orbits and span a wide range of initial eccentricities and inclinations. The author finds that the stability criteria derived from their SVM model works significantly better than previous criteria for mutual inclinations less than 40$^\circ$.


\cite{vynatheyaAlgebraicMachineLearning2022a} used multilayer perceptron (MLP) models to classify triple-star systems as stable and unstable. The models were trained on results of a large ensemble of N-body simulations ($10^6$ systems). They focus on moderate mass ratio ($10^{-2} {\textless}\mu_{1} {\textless} 1, 10^{-2} \textless \mu_{2} {\textless} 10^2$) systems, with initial conditions spanning the entire range of inner and outer eccentricities and mutual inclinations. The trained model showed better agreement with N-body results (with 95\% accuracy) as compared to algebraic criteria.

\section{Conclusion}
The study of planetary system stability, especially in compact multi-planetary and hierarchical systems, has seen significant advancements in recent decades. Our understanding of the intricate dynamics governing these systems has deepened through a combination of analytical theories, N-body simulations, and cutting-edge machine learning techniques.

In compact multi-planetary systems, the overlap of mean motion resonances (MMRs) between adjacent planetary pairs and three-body resonances have been identified as critical factors influencing stability. Theoretical models and extensive numerical simulations have provided insights into the necessary planetary separations, typically expressed in terms of mutual Hill radii, to maintain long-term stability. These theoretical predictions align well with observed exoplanetary systems, shedding light on the processes shaping planetary architecture and formation mechanisms. This allow estimation on the stability of planetary systems much more efficiently. 

Hierarchical systems, generally considered more stable, exhibit complex dynamical behaviors when influenced by secular resonances, particularly in systems with highly inclined or eccentric orbits. As orbits become highly eccentric, the potential for instability increases. Advances in the secular theory of hierarchical dynamics and the derivation of stability limits have enhanced our ability to predict and understand these instabilities. 

Recently, machine learning approaches have been successfully employed to assess stability in both compact and hierarchical systems, providing efficient and accurate predictions. Overall, the integration of traditional analytical methods with modern computational techniques has significantly advanced our understanding of planetary system stability. This comprehensive approach aids in the characterization of exoplanetary systems.

\bibliographystyle{Harvard}
\bibliography{reference,refStability,hierarchicalMultiples,threebodydynamics}
\end{document}